\documentstyle[epsfig]{mn}
 at 10truept
\title
     [ Estimating Cosmological Parameter Covariance]
{\vglue-3.0truecm \centerline{\it\small 
For submission to Monthly Notices}
\vglue 2.5truecm
 Estimating Cosmological Parameter Covariance
\author
     [ Andy Taylor, Benjamin Joachimi]
     { Andy Taylor$^1$\thanks{ant@roe.ac.uk} \& Benjamin Joachimi$^{1,2}$\\
 	$^1$ Scottish Universities Physics Alliance, 
     Institute for Astronomy,
     School of Physics and Astronomy,
     University of Edinburgh,\\
     \,\,\, Royal Observatory,
     Blackford Hill,
     Edinburgh, EH9 3HJ,
     U.K.\\
     $^2$ Department of Physics and Astronomy,  University College London, Gower Street, London, WC1E 6BT, UK}}
\newcommand{\be}{\begin{equation}}
\newcommand{\ee}{\end{equation}}
\newcommand{\ba}{\begin{eqnarray}}
\newcommand{\ea}{\end{eqnarray}}
\newcommand{\nn}{\nonumber \\}

\newcommand{\thetab}{\mbox{\boldmath $\theta$}}

\newcommand{\de}{\partial}

\newcommand{\lgl}{\langle}
\newcommand{\rgl}{\rangle}

\newcommand{\Tr}{\mbox{\rm Tr}}
\newcommand{\lb}{\mbox{\boldmath $\ell$}}

\newcommand{\Lb}{\mbox{\boldmath $L$}}

\newcommand{\D}{\mbox{\boldmath $D$}}
\newcommand{\C}{\mbox{\boldmath $M$}}   

\newcommand{\I}{\mbox{\boldmath $I$}}
\newcommand{\A}{\mbox{\boldmath $A$}}

\newcommand{\V}{\mbox{\boldmath $V$}}

\newcommand{\calP}{\mbox{$\mathcal{P}$}}
\newcommand{\calL}{\mbox{$\mathcal{L}$}}
\newcommand{\Mb}{\mbox{\boldmath $M$}}
\newcommand{\mub}{\mbox{\boldmath $\mu$}}

\newcommand{\calF}{\mbox{$\mathcal{F}$}}

\newcommand{\calC}{\mbox{$C$}}

\newcommand{\Psib}{\mbox{\boldmath $\Psi$}}
\newcommand{\U}{\mbox{\boldmath $U$}}
\newcommand{\B}{\mbox{\boldmath $B$}}
\newcommand{\Sigmab}{\mbox{\boldmath $\Sigma$}}
\newcommand{\R}{\mbox{\boldmath $R$}}

\begin{document}

\maketitle

\begin{abstract} 
We investigate the bias and error in estimates of the cosmological parameter covariance matrix, due to sampling or modelling the data covariance matrix, for likelihood width and peak scatter estimators.  We show that  these estimators do not coincide unless the data covariance is exactly known. 
For sampled data covariances, with Gaussian distributed data and parameters, the parameter covariance matrix estimated from the width of the likelihood has a Wishart distribution,  from which we derive the mean and covariance.
This mean is biased and we propose an unbiased estimator of the parameter covariance matrix. 
Comparing our analytic results  to a numerical Wishart sampler of the data covariance matrix we find excellent agreement. 
An accurate {\it ansatz} for the mean parameter covariance for the peak scatter estimator is found,  and we fit  its covariance to our numerical analysis. 
The mean is again biased and we propose an unbiased estimator for the peak parameter covariance. 
For sampled data covariances the width estimator is more accurate than the peak scatter estimator. 
We  investigate modelling the data covariance, or equivalently data compression, and shown that the peak scatter estimator is less sensitive to biases in the model data covariance matrix than the width estimator, but requires independent realisations of the data to reduce the statistical error.  If the model bias on the peak estimator is sufficiently low this is promising, otherwise the sampled width estimator is preferable. 
\end{abstract}


\section{Introduction}


The high precision required to probe the nature of dark energy, dark matter and modifications to gravity (e.g., Amendola et al. 2013) is driving cosmology to an era where the accuracy of parameter estimation will have to reach sub-percent levels.
To meet this challenge  large-scale ground and space-based cosmological surveys are being planned and carried out which are optimised to deliver high statistical accuracy (e.g., VST-KiDS, DES, HSC, LSST, Euclid). 
For these surveys to be successful systematic biases will also have to be controlled to an unprecedented level, within the bounds set by the statistical uncertainty. The introduction of statistical uncertainty and systematic biases will have to be tracked at every step of the data analysis,  from observation to parameter estimation.


An aspect which has recently been receiving more attention in this process is  the final parameter estimation step when data, compressed into the form of power spectra or correlation functions, is compared with cosmological models and further compressed into estimates of the model parameters along with an estimate of their accuracy. In particular we need to have reliable, unbiased estimates of the  parameter covariance matrix, which is needed to demonstrate how accurate the parameters have been measured as well as delineating the volumes of parameter space where acceptable models reside. 
Beyond this, if we want to apply some form of model selection, for example investigating the Bayesian Evidence, we need to have an accurate representation of the posterior distribution of the parameters. If the parameter covariance matrix is biased by a poor estimator it will either over- or underestimate the actual errors and covariances of the measured parameters. In addition, a sub-optimal covariance estimator will itself have significant uncertainties which should be folded into the overall error budget.


There are two common  approaches to estimating the uncertainty on parameters derived  from cosmological data. One is to estimate the variance, or width, of the likelihood surface in parameter space. This can be done by mapping out the likelihood surface and numerically integrating on a grid,  or using a Monte-Carlo Markov Chain (e.g., Lewis \& Bridle 2002) to sample the likelihood distribution and Monte-Carlo integrating.  A second approach is to generate   many independent  realisations of the survey, either by simulating or re-sampling the data, estimate the maximum likelihood parameter values for each realisation, and then use the scatter in the peak values as an estimate of the uncertainty in the result. In the limit of no bias or uncertainty in the model distribution, these should yield the same answer. However, as we shall show, if the likelihood distribution does not accurately model the distribution of the data both of these estimators will produce biased estimates of the parameter covariance matrix, and the width and peak scatter estimators will no longer coincide. In addition, biases in the likelihood function can significantly increase the uncertainty in the parameter covariance to the point that the error estimate is unreliable.

For data which follows a multivariate Gaussian distribution all of the statistical information is encoded within  the data covariance matrix and in particular its inverse, the precision matrix. 
If the data covariance is known exactly {\em a priori}, it can be accurately inverted to find the precision matrix and the likelihood function is unbiased.  However, if the data covariance is not well known, or must be estimated, it will  be biased and this bias will propagate into the likelihood function through the precision matrix.

 In Cosmology, where  the nonlinear evolution of density perturbations and effect of baryons, galaxy formation, stellar and AGN feedback affect the statistical properties of observables in a complex way (e.g., Semboloni et al. 2011, van Daalen et al. 2013), the data covariance matrix is usually  estimated by sampling independent realisations of the data, either from simulations or re-sampling the data using Jackknife (e.g. Tukey 1958) or Bootstrap (e.g., Efron 1979, Norberg et al. 2009) methods, or from models of the data covariance matrix which aim to contain nonlinearity and feedback effects  (e.g., Cooray \& Hu 2001, Takada \& Bridle 2007, Takada \& Jain 2009, Hilbert et al. 2011, Kayo et al. 2012). 
Estimating the data covariance by sampling  independent realisations of the data will introduce a sampling variance which propagates into the precision matrix and parameter estimation. 
For Gaussian-distributed data the sample data covariance matrix follows a Wishart distributed (Wishart 1928), while the precision matrix follows an Inverse-Wishart distribution, first studied by Kaufmann (1967; see also, e.g., Taylor et al. 2013). Kaufmann derived the expectation value of the precision matrix and showed it was biased compared to the inverse of the expected data covariance, and that this bias diverges when the number of realisations approaches the size of the data set. Hartlap et al. (2007)  showed that this bias is found in cosmological precision matrices when the data covariance matrix is estimated from an ensemble of cosmological simulations, and suggested that the bias should be corrected for to avoid underestimating parameter errors.

Taylor et al. (2013) extended this analysis to estimating cosmological parameter covariance matrices and, assuming that the parameters were Gaussian-distributed in parameter space, showed to second order that the variance, or width, of the likelihood function would have an additional  uncertainty due to the sampled precision matrix. 
Dodelson \& Schneider (2013)  further showed that if the parameter covariance was estimated from the scatter in the peak of the likelihood the parameter covariance would be biased high, to second order, due to the Inverse-Wishart scatter of the precision matrix. 
Likewise, Percival et al. (2014) showed a similar bias arises to second order in the width estimator.

In this paper we complete this analysis  by studying the distribution, bias and covariance of the parameter covariance matrix from the width and peak scatter estimators when the data covariance matrix is sampled from independent realisations of the data.
In addition we investigate the effect of modelling the data covariance matrix,  where there is no sampling variance, but uncorrected biases will propagate into the parameter covariance matrix.  In Section 3 we discuss how to numerically  generate random realisations of the data covariance matrix to compare with our results. In Section 4 we derive the exact distribution for the parameter covariance matrix estimated from the width of the likelihood, and its bias and error. We also use our numerical results to propose an ansatz for the bias in the peak scatter estimator and a fit to its error. We study modelling of the data covariance and data compression in Section 5, and present our summary and conclusions in Section 6. We begin by reviewing estimators for the parameter covariance and how a bias in the data covariance matrix propagates.

\section{Parameter covariance}

For a given set of data, $\D$, cosmological parameters, $\thetab$, can be estimated by sampling the posterior parameter distribution, 
$\calP(\thetab|\D)\propto \calP(\D|\thetab)\calP(\thetab)$,
where the likelihood distribution of the data is $\calP(\D|\thetab)$,  and the parameter prior is  $\calP(\thetab)$. 
We will  focus on the case where the data follows  a Gaussian distribution and the mean of the likelihood  depends on the cosmological parameters, $\mub(\thetab)$, while the data covariance matrix, $\C = \lgl \Delta \D \Delta \D^t \rgl$, is independent of the parameters,  and $\Delta \D = \D - \lgl \D \rgl$ is the fluctuation of the data around estimates of the mean. The log-likelihood is given by $\calL=-2 \ln \calP(\D|\thetab) = \Delta \D^t {\Psib} \Delta \D$, where $\Psib = \C^{-1}$ is the inverse data covariance matrix, the precision matrix.

If the data covariance matrix is estimated with some uncertainty  we can treat it, and the precision matrix, as random variables and marginalise over the uncertainty in the likelihood function with a prior on the precision matrix, 
\be
	\calP (\thetab| \D, \Psib) = \int  \! d\widehat{\Psib} \,  \calP(\thetab | \D, \widehat{\Psib})\, \calP(\widehat{\Psib}|\Psib),
\ee
where $\Psib$ is the true precision matrix, $\widehat{\Psib}$ is its estimated value and $\calP(\widehat{\Psib}|\Psib)$ is the prior.
In the case that the precision matrix is known the prior will be a delta-function. But if the mean of the data and precision matrix is estimated from the inverse of the sampled data covariance matrix, 
 \be
 	\widehat{\Psib} =\widehat{\C}^{-1}= \left[ \frac{1}{N_S-1}  \sum_{i=1}^{N_S} \Delta \D_i \Delta \D^t_i \right]^{-1},
\ee
 where $\Delta \D_i$ is the $i^{\rm th}$ realisation from $N_S$ random Gaussian samples, the prior is  Inverse-Wishart distributed\footnote{
Sometimes the Inverse-Wishart prior is modified by swapping the roles of  $\widehat{\Psib}$ and $\Psib$ in the prior to make the integration tractable for Gaussian-distributed parameters. This is called a ``Natural Conjugate Prior" (see, e.g., Press 1982). However, while analytically useful, it is unjustified and we do not consider this further.
}. 
If the data covariance matrix is an analytic model the prior may also be Inverse Wishart if we assume random inaccuracies.

We are interested in any biases in the parameter covariance matrix. The $n$-th order moments of the parameter distribution are found from integrating over the parameter distribution,
 \be
 	\lgl \Delta \theta_{\alpha_1} \cdots \, \Delta \theta_{\alpha_n} \rgl_\theta = \int \! d^{\small N_{\!P}} \! \theta \,  \Delta \theta_{\alpha_1}
	\cdots \, \Delta\theta_{\alpha_n}
	\calP(\thetab|\D,\Psib).
 \ee
 In particular, the covariance matrix of the parameter is $C_{\alpha\beta} = \lgl \Delta \theta_\alpha \Delta \theta_\beta \rgl $, where $\Delta \theta_\alpha = \theta_\alpha - \lgl \theta_\alpha \rgl $ is the off-set from the mean.
 Marginalising over the precision matrix and calculating the moments commute, so we can estimate  moments of the posterior for a fixed, sampled precision matrix and then marginalise over the precision prior, $\calP(\widehat{\Psib}|\Psib)$.
 
If we assume the posterior  is also Gaussian distributed the log-likelihood can be expanded to second order in parameter space\footnote{Here and in the following we employ the Einstein sum convention.}, $\calL=\calL_0 + \Delta \theta_\alpha \calL_\alpha+ \Delta \theta_\alpha \Delta \theta_\beta \calL_{\alpha\beta}/2$, where $\calL_\alpha$ is the gradient of the log-likelihood with respect to the parameters and $\calL_{\alpha\beta}$ is the curvature. We can approximate the curvature with its expectation with respect to the data, $\lgl \calL_{\alpha\beta} \rgl= 2 \calF_{\alpha\beta}$, where $\calF_{\alpha\beta} = \A_\alpha^t \Psib \A_\beta$ is the Fisher Information matrix (e.g., Tegmark, Taylor \& Heavens 1997) and $\A_\alpha = \de_\alpha \mub$ is the gradient of the model in parameter space. 
Assuming a flat prior on the parameters, the mean of the parameter distribution is 
 \be
 	\lgl \Delta \theta_\alpha \rgl_\theta = - \calF^{-1} _{\alpha\beta} \, d_\beta,
	  \label{eq:peak}
 \ee
 which coincides with the maximum for a Gaussian distribution. As $d_\alpha = \A^t_\alpha \Psib \Delta \D$ vanishes when averaged over the data, $\lgl d_\alpha \rgl_D = 0 $, the mean is always an unbiased estimate of the peak of the likelihood.  The covariance of the parameters around the peak  is
 \be
 	\lgl \Delta \theta_\alpha \Delta \theta_\beta \rgl = 
	\calF_{\alpha\beta}^{-1}  
	,	
 \ee
which is a measure of the width and shape of the likelihood surface.  
Alternatively, we can create an ensemble of realisations of the data and estimate the peak likelihood found from equation (\ref{eq:peak}). The scatter in the peaks is $\lgl \Delta \theta_\alpha \Delta \theta_\beta \rgl_\theta =   \calF_{\alpha\mu}^{-1}  \lgl d_\mu d_\nu \rgl \calF_{\nu\beta}^{-1}=\calF_{\alpha\beta}^{-1} $. Hence,  when the data covariance matrix is accurately known the width of the likelihood and scatter in the peak values coincide.



 If we vary the Fisher matrix from its true value the parameter covariance estimated from the width of the likelihood becomes 
$
	\widehat{\calC}^W_{\alpha\beta} = (\calF + \Delta \calF)^{-1}_{\alpha\beta} ,
$	
 and the change in the parameter covariance is
 \be
 	\Delta \calC^W_{\alpha\beta} =  - \calF^{-1}_{\alpha\mu} (I_{\mu\nu'} + \Delta \calF_{\mu\mu'} \calF^{-1}_{\mu'\nu'})^{-1} \Delta \calF_{\nu'\nu}  \calF^{-1}_{\nu\beta}.
	\label{eq:curv-cov}
\ee
 We can expand this expression to second order,  $\Delta \calC^W = - \calF^{-1} [\Delta \calF  -   \Delta \calF  \calF^{-1} \Delta \calF] \calF^{-1}$  where $\Delta \calF \ll \calF$. If the change in the Fisher matrix is due to a change in the precision matrix,  $\Delta \calF_{\alpha\beta} =\A_\alpha^t \Delta \Psib \A_\beta$,  this propagates through to the width estimator to first order.

 
 We can also estimate a similar bias in the parameter covariance matrix from  the scatter in the maximum likelihood.  If we vary the sample precision matrix, $\widehat{\Psib}$, so that $\widehat{\Psib} = \Psib + \Delta \Psib$, the estimated maximum likelihood is
 \be
 \Delta\widehat{ \theta}_\alpha = - [\A^t_\alpha (\Psib+\Delta \Psib) \A_\beta ]^{-1} \A_\beta^t  (\Psib + \Delta \Psib) \Delta \D.
 \label{eq:peak_change}
 \ee
 The expectation value remains zero, $\lgl \Delta \widehat{\theta}_\alpha \rgl =0 $, and  the parameter estimator is again unbiased. 
 The covariance of this  is
 $
 	\widehat{C}^P_{\alpha\beta } = \lgl \Delta \widehat{\theta}_\alpha \Delta \widehat{\theta}_\beta \rgl = 
		\widehat{\calF}_{\alpha\mu}^{-1} \lgl \widehat{d}_\mu \widehat{d}_\nu \rgl  \widehat{\calF}_{\nu \beta}^{-1} 
 $
where $\widehat{\calF}=\calF + \Delta \calF$ and $\widehat{d} = d + \Delta d$. If the precision matrix is equal to its expected value this reduces to $\calC = \calF^{-1}$, and the scatter in the likelihood peak is an unbiased estimator of the parameter covariance matrix. 
We can write the change in the parameter covariance as
\be
		\Delta{C}^P_{\alpha\beta }=  (\calF + \Delta \calF)^{-1} _{\alpha\mu}
			[\Delta \Delta \calF - \Delta \calF \calF^{-1}\Delta \calF]_{\mu\nu}
		 (\calF + \Delta \calF)^{-1} _{\nu\beta},
		 \label{eq:exact_cov}
\ee
where  $ \lgl d_\alpha \Delta d_\beta  \rgl  = \lgl \Delta d_\alpha d_\beta\rgl = \Delta \calF_{\alpha\beta} $  and
 $\Delta \Delta \calF_{\alpha\beta} = \lgl  \Delta d_\alpha   \Delta d_\beta \rgl =\A_\alpha^t ( \Delta \! \Psib \, \C \, \Delta \!\Psib) \A_\beta$.  There are two competing effects here, the change in the likelihood curvature and the gradient.  
   An increase in the precision matrix will increase the curvature and decrease the parameter covariance, while the  scatter in the peak of the likelihood will increase the parameter covariance. 
 These two effects cancel to first order in the peak parameter covariance, indicating that the peaks of the likelihood are less sensitive to changes in the precision matrix than the width. If the variation in the precision matrix is proportional to the precision matrix then $\Delta \Delta \calF = \Delta \calF \calF^{-1} \Delta \calF$ and the change in the peak covariance vanishes. This is due to the cancellation of the normalisation of the precision matrix in the peak estimator equation (\ref{eq:peak_change}). Any other dependence of the precision matrix will depend on the combination $\A_\alpha^t \Delta \Psib$ and only arises to second order and higher. 
 Another interesting cancellation occurs when $N_D=N_P$.
 In this case the response matrix, $\A_\alpha$, is square with dimensions $N_P \times N_P$ and is invertible unless singular. Assuming 
 the inverse of the Fisher matrix can be written $\calF_{\mu\nu}^{-1}= \A^{-1}_\mu \C (\A^t)^{-1}_{\nu}$  we find  that $\Delta \calF \calF^{-1} \Delta \calF = \Delta \Delta \calF$ and again  equation $\Delta {C}^P_{\alpha\beta}$ vanishes. We explore this further in Section \ref{sec:data_compress}.

 \section{A Monte-Carlo Wishart Sampler}
 
\subsection{Generating Wishart random matrices}

To test their results,  Taylor et al. (2013) and Dodelson \& Schneider (2013) created large numbers of simulated realisations of data sets  to determine the sample properties of the covariance and precision matrices. Here, we take a computationally more efficient route by directly generating random realisations of the data covariance from a Wishart distribution. Odell \& Feiveson (1966) proposed a simple routine to produce samples of arbitrary Wishart-distributed matrices using only independent, univariate samples from a Gaussian Normal distribution  and a $\chi^2$-distribution. Their method is based on the Bartlett decomposition for a Wishart-distributed matrix, 
$
	{\C} = \Lb {\U}{\U}^t \Lb^t,
$
where $\lgl \C \rgl=\Lb \Lb^t$ is the Cholesky decomposition of the expectation of ${\C}$, and  ${\U}{\U}^t$ is the Bartlett decomposition of a Wishart-distributed identity matrix. The elements of ${\U}$ are distributed as
$
	U_{ii}^2 \sim \chi^2(\nu+i-1), $ 
 $	U_{ij} \sim {\cal N}(0,1) ~{\rm for}~ i>j, $ 
  and $	U_{ij} = 0 ~{\rm for}~ i<j, 
$
where $``\sim"$ means  ``is drawn from the distribution", and where $\chi^2(n)$ denotes a $\chi^2$-distribution with $n$ degrees of freedom.
Hence, ${\U}$ is a lower triangular matrix with $\chi^2$-distributed diagonals and standard normal-distributed off-diagonals. To calculate the means and variances we use 1000 samples when considering matrix traces, and increase this number to $10^5$ when discussing statistics of individual matrix elements.

\subsection{Modelling cosmological data}

For all numerical calculations we assume a generic weak gravitational lensing survey to produce the underlying data set, constraining the parameters of a standard flat $w$CDM cosmology. As shown by Taylor et al. (2013), the scalings of means and variances of data and parameter covariances are insensitive to the specific  data and parameter set under consideration. Hence, we expect our conclusions not to depend on the details of the model.

We assume a data vector composed of weak gravitational lensing convergence power spectra, $P_\kappa(\ell)$, measured in $N_\ell=24$ angular frequency bins, logarithmically spaced between $\ell=50$ and $\ell=5000$. The underlying matter power spectrum is calculated using the transfer function by Eisenstein \& Hu (1999), and non-linear corrections according to {halofit} (Smith et al. 2003). The redshift distribution of sources follows the scaling
\be
	p(z) \propto z^2 \exp \left\{ - \left( \frac{z}{z_0}\right)^{1.5} \right\}\;,
\ee
where $z_0 =0.64 $, which corresponds to a median redshift of 0.9. The distribution is truncated below $z=0.2$ and above $z=2$. The uncertainty induced by photometric redshift estimates is modelled as a Gaussian scatter of width $\sigma_z=0.05(1+z)$ around the true redshift. 
To compute the data covariance, we assume that the convergence is Gaussian distributed, so that the power spectrum covariance is given by, e.g., equation (53) of Joachimi et al. (2008). We set the mean galaxy number density to $n_{\rm g}=30\,{\rm arcmin}^{-2}$, the intrinsic ellipticity dispersion, or shape noise, to $\sigma_\epsilon=0.35$, and the survey area to $A_{\rm s}=15,000\,{\rm deg}^2$. We also consider a lognormal covariance model which is described in Section~\ref{sec:data_compress}.
The default cosmological parameter set has size $N_P=7$ and consists of the matter and baryon density parameters $\Omega_{\rm m}=0.3$, $\Omega_{\rm b}=0.045$, the Hubble parameter $h=0.7$, the slope of the primordial matter power spectrum $n_{\rm s}=1$, the normalisation of matter density fluctuations $\sigma_8=0.8$, and the dark energy equation of state parameters $w_0=-1$ and $w_a=0$.

\begin{figure}
\centering
\includegraphics[scale=1]{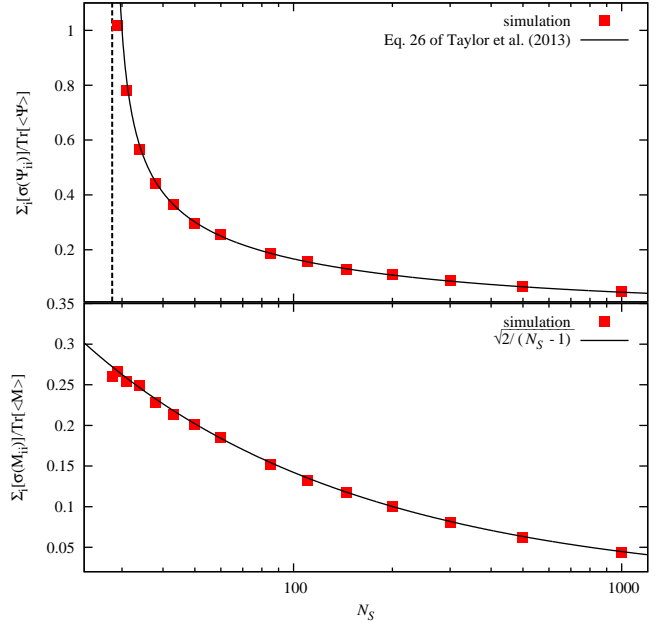}
\caption{\textit{Top panel}: Sum of the standard deviation  of the diagonal elements of the inverse data covariance, the precision matrix, as a function of the number of realisations used to generate the data covariance, $N_S$. The standard deviation is normalised by the trace of the noise-free precision matrix. Red squares correspond to Wishart-sampled simulation results, the black solid line to the Inverse-Wishart scaling given by Kaufmann (1967). The vertical black dashed line marks the divergence of this scaling. \textit{Bottom panel}: The trace of the variance of the data covariance matrix. Red squares are from the Wishart-sampler, while the solid line is the expected scaling.}
\label{fig:datacov}
\end{figure}

 \section{Sampling estimators }
 
\subsection{The data covariance and precision matrix}

Estimates of cosmological data covariance matrices and  precision matrices can be difficult to make, due to the complex combination of nonlinear evolution of the density field,  the effects of baryons and feedback in galaxy formation where the physical processes are not yet well understood. While much effort has gone into analytic modelling of the data covariance matrix (e.g., Cooray \& Hu 2001, Takada \& Bridle 2007, Takada \& Jain 2009, Hilbert et al. 2011, Kayo et al. 2012), a more straightforward, although computationally expensive, approach is to numerically simulate a large volume of the Universe to model the survey. This approach allows us to generate independent, random realisations of the survey given a cosmological model. Each realisation is  analysed using the same analysis pipeline as applied to the real data, and the statistical properties of the results studied. While this is highly versatile, one drawback is that each estimate of the data covariance matrix is a random sample of the model data covariance matrix. If the data is Gaussian-distributed and we have $N_S$  realisations of the survey, the data sample covariance matrix is Wishart-distributed with $N_S-1$ degrees of freedom, $\widehat{\C}\sim W_{N_D}(\C,N_S-1)$, where we assume throughout that the mean of the data covariance matrix is estimated from the data\footnote{We define $W_p(\Mb,n) = P(\widehat{\Mb}|\Mb,p,n)$ in the notation of Taylor et al. (2013), where $\Mb$ is a $p \times p $ matrix with $n$ degrees of freedom.} (see, e.g., Taylor et al. 2013). The precision matrix is Inverse-Wishart  distributed with $N_S-N_D-2$ degrees of freedom, where $N_D$ is the number of data-points in the analysis, $\widehat{\Psib}\sim W^{-1}_{N_D}(\Psib,N_S-N_D-2)$. Contrary to what we would expect from Gaussian or Wishart statistics, the statistical properties of the precision matrix depend not only on the number of samples, $N_S$, but also the size of each sample, $N_D$. This arises due to the change in variable transforming from the data covariance to the precision matrix. The sampling variance in the estimated precision matrix will then propagate into the parameter covariance. The expectation of the precision matrix from an ensemble of realisations, 
$
	\lgl \widehat{\Psib}\rgl = [(N_S-1)/(N_S-N_D-2)] \Psib
$
(Kaufmann 1967, Hartlap et al. 2007), is biased but can be corrected. The covariance of the precision matrix is given by equation (26) of  Taylor et al. (2013). In Figure  \ref{fig:datacov} we plot the predicted sum of the variance of diagonals of the precision matrix and the results from our Wishart-sampler, as well as the variance of the data covariance matrix. In both cases we find the Wishart sampler and the analytic prediction agree very well. 

Hartlap et al. (2007) have shown the predicted bias in the precision matrix also holds in cosmological N-body simulations of weak lensing, while Dodelson \& Schneider (2013) have shown the same is true for the bias in the parameter covariance estimated from peak scatter. This suggests that the scalings derived in this work will remain valid for data sets whose distribution clearly departs from Gaussianity.

\subsection{Parameter covariance from the likelihood width}

\begin{figure}
\centering
\includegraphics[scale=1.]{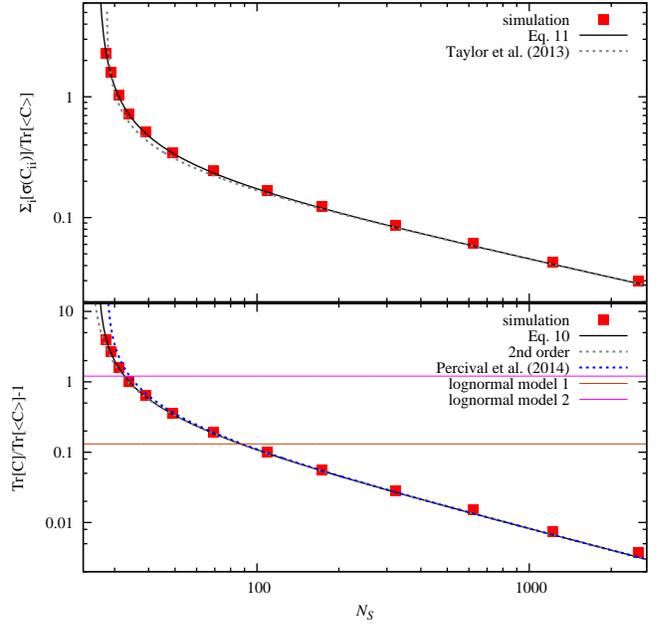}
\caption{\textit{Top panel}: Sum of the standard deviations of the diagonal elements of the parameter covariance matrix measured from the likelihood width as a function of the number of realisations used to generate the data covariance matrix, $N_S$. The standard deviation is normalised by the trace of the noise-free parameter covariance. Red squares correspond to Wishart-sampler simulation results, the black solid line to the scaling given by equation (\ref{eq:widthcov}). The grey dotted line is obtained from the second order prediction of Taylor et al. (2013). \textit{Bottom panel}: Relative deviation of the trace of the mean parameter covariance from its expectation as a function of $N_S$. Red squares again show Wishart-sampled simulation results while the black solid line is the scaling given by equation (\ref{eq:DCC}). The black dotted line is a second order approximation, while the blue dotted line is the solution given by Percival et  al. (2014).
The horizontal pink (orange) line corresponds to the bias caused by the use of the lognormal model covariance in Model 1 (2) in Section \ref{sec:data_compress}.}
\label{fig:parcov_curv_nd-24}
\end{figure}

Since the parameter covariance estimated from the width of the posterior distribution is the inverse of the Fisher matrix, which itself is a linear transformation of the precision matrix, we might expect the parameter covariance matrix to be  Wishart-distributed. This is indeed the case, as we show here.
Suppose $\V$ is a nonsingular, $p \times p$ symmetric Wishart-distributed matrix with $n$ degrees of freedom, $\V \sim W_p(\Sigmab,n)$, where $\Sigmab$ is the expectation value of $\V$,  and $\B$ is a $r \times p $ matrix of rank $r$, then the matrix $(\B \V^{-1} \B^t)^{-1}\sim W_r( [\B \Sigmab^{-1} \B^t]^{-1},n-p+r ) $ is also Wishart distributed (Eaton 2007). Hence
 when both the data and the parameters are Gaussian-distributed, and the data covariance matrix is estimated from an ensemble of independent realisations of the data,  the parameter covariance matrix has the Wishart distribution   
 \be
	 \widehat{\calC}_{\alpha\beta}=\widehat{\calF}^{-1}_{\alpha\beta} \sim W_{N_P}(\calC_{\alpha\beta},N_S-N_D+N_P-1).
 \ee
  The degrees of freedom, $N_S-N_D+N_P-1$, reflect both the behaviour of the precision matrix which introduces the size of the data-set, $N_D$, and the compression of information into the $N_P$ cosmological parameters. From this distribution we find the expectation value  of the parameter covariance matrix is
  \be
 	\left\lgl \widehat{\calC}^W_{\alpha\beta} \right\rgl 
		= \frac{(N_S-N_D+N_P-1)}{(N_S-N_D-2)} \, \calC_{\alpha\beta}.
			\label{eq:DCC}
\ee
 The bias in the parameter covariance for the width estimator depends only on the difference between number of sampled realisations and data size,  $N_S-N_D$, and number of parameters $N_P-1$. This bias diverges at $N_S-N_D=2$, as the data covariance is formally uninvertible, while for $N_D=N_P$  the pre-factor reduces to $(N_S-1)/(N_S-N_D-2)$, the factor we divided the precision matrix by to correct for its bias. In this case the step of inverting the data covariance matrix was unnecessary as the parameter covariance matrix can be written directly in terms of the data covariance matrix.
  Expanding equation (\ref{eq:DCC}) when $N_S-N_D \gg N_P$  we find the extra covariance is $\lgl \Delta \calC^W_{\alpha\beta} \rgl \approx (N_P+1)/(N_S-N_D)[1+2/(N_S-N_D)]$, where the second term slightly differs from the result of Percival et al. (2014) due to the different order of expansion. In the limit that  $N_S \gg N_D \gg N_P \gg 1$  the extra covariance is $\lgl \Delta \calC^W_{\alpha\beta} \rgl \approx (N_P/N_S) \,\calC_{\alpha\beta} $, showing that an increase in the number of model parameters must be compensated for by an increase in the number of realisations.  
 The bottom panel of Figure \ref{fig:parcov_curv_nd-24} shows the predicted and Wishart-sampled bias for the width estimated parameter covariance as a function of realisations, $N_S$, for $N_D=50$ and $N_P=6$, as well as  the second order predictions  which appeared in this paper and Percival et al. (2014). We find that  our exact analytic results agree very well with the Wishart-sampled simulations.

\begin{figure}
\centering
\includegraphics[angle=-90,scale=0.425]{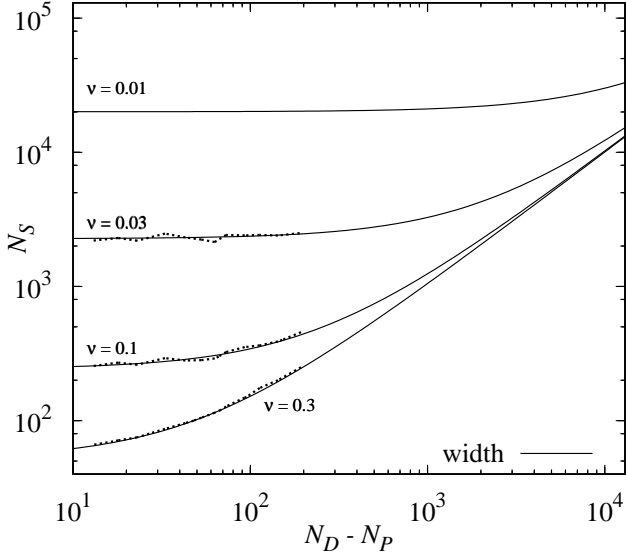}
\vspace{-0.1cm}
\caption{The fractional error, $\nu$, from the combined bias and variance on the parameter covariance matrix as a function of  the number of realisations used to estimate the data covariance matrix, $N_S$, and  the size of the data vector minus the number of parameters fitted, $N_D-N_P$. The black solid lines  show  different values of constant fractional error for the width parameter covariance estimator. For $N_D < 200$ the corresponding results from the Wishart-sampled simulations are overplotted as dotted lines, demonstrating excellent agreement.
}
\label{fig:fom}
\end{figure}

 As well as the bias in the covariance it is worth estimating the uncertainty in our estimate of the parameter covariance matrix, since  this can become the dominant source of error. The covariance of the parameter covariance from the Wishart distribution is 
 \be
 	\!\left\lgl \Delta \calC^W_{\alpha\beta} \Delta \calC^W_{\mu\nu} \right\rgl 
	\! = \! \frac{N_S\!-\!N_D\!+\!N_P\!-\!1}{(N_S-N_D-2)^2} \!
	\left( \calC_{\alpha\mu} \calC_{\beta\nu} \! + \! \calC_{\alpha\nu} \calC_{\beta\mu} \right). 
	\label{eq:widthcov}
 \ee
 The covariance of the parameter covariance diverges again at $N_S-N_D=2$, while for $N_S=N_D$ the covariance reduces to the bias correction to the precision matrix and the factor $1/(N_S-1)$, which is just the usual scaling of the covariance matrix for independent Gaussian realisations. The covariance has a similar form to Wick's Theorem for a Gaussian variable but the coefficient is significantly different, and a  Gaussian approximation to the distribution of the parameter covariance matrix would overestimate the variance by a factor equal to the number of degrees of freedom.
 The error on a parameter variance 
		for $N_S-N_D \gg N_P-1$ reduces to $\sigma[\calC^W_{\alpha\alpha}] = \sqrt{2/(N_S-N_D)}\calC_{\alpha\beta}$, in agreement with the second order result found by Taylor et al. (2013).		  
	The upper panel of Figure \ref{fig:parcov_curv_nd-24} shows the exact analytic and Wishart-sampled sum of the variance of the diagonal terms of the parameter covariance, which agree very well. We also plot the second order result from Taylor et al. (2013) which deviates from both.

Combining in quadrature the predicted bias in the parameter covariance (equation \ref{eq:DCC}) with the variance of the covariance, $\lgl |\Delta\calC_{\alpha\beta}^W|^2\rgl $ from equation (\ref{eq:widthcov}),  and defining $\nu=\Delta{\calC}/\calC $ as the fractional increase in the covariance, we solve to find the number of realisations needed to reach a given accuracy for a given data set and model;
 \be
 	N_S = N_D +2 + \nu^{-2}[1 + \sqrt{1+\nu^2 (N_P+1)(N_P+3)}] ,
	\label{eq:NSwidth}
\ee
which can be approximated for simplicity to $N_S \approx   N_D +N_P/\nu+ 2 \nu^{-2}$.
 If we require an accuracy of $\nu=0.1$ on the parameter covariance matrix we need $N_S \approx N_D+ 10N_P +200$ realisations. When $\nu N_P \ll 2$ this agrees with the earlier result of Taylor et al. (2013).  Figure \ref{fig:fom} shows contours of the fractional error on the parameter covariance matrix from this combined bias and error. We also show for smaller $N_S$ the agreement between our analytic expectation and the Wishart sampler. For large values of $N_D-N_P$, all of the lines asymptote to $N_S = N_D$.

\begin{figure}
\centering
\includegraphics[scale=0.7]{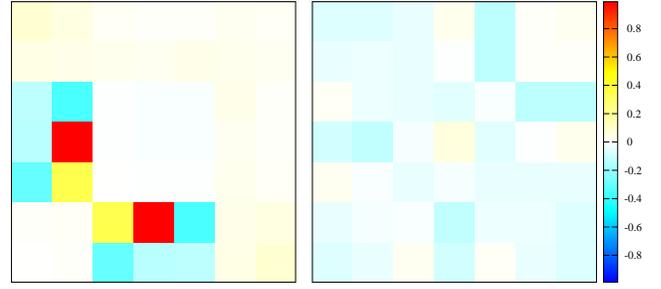}
\caption{\textit{Left panel}: Percentage difference  between Wishart sampled simulation and analytic expression for the mean of the elements in the parameter covariance matrix determined from the width of the likelihood using $N_S=100$ realisations. \textit{Right panel}: Same as the left panel, but for the relative difference in the standard deviation of the parameter covariance.}
\label{fig:fullmatrix_curv}
\end{figure}

As well as the diagonal terms in the parameter covariance matrix, we would also like to test the off-diagonal elements. The left panel in Figure \ref{fig:fullmatrix_curv} shows the relative difference between the values of the expected 7-parameter covariance matrix and the values from the Wishart-sampled simulation with $N_S=100$. All deviations are below  $2\%$ and most are less than $0.2\%$. The largest deviations appear in the estimation of the correlation  between $\sigma_8$ and $n_{\rm s}$, but is at a low enough level to be simply due to statistical fluctuations in the Wishart-sampler. The right panel of Figure \ref{fig:fullmatrix_curv}  shows the relative deviation between the analytic result for the variance of the parameter covariance matrix and the variance found in the simulated ones. Again, the fractional deviation is within a fraction of a percent.

\begin{figure}
\centering
\includegraphics[scale=0.85]{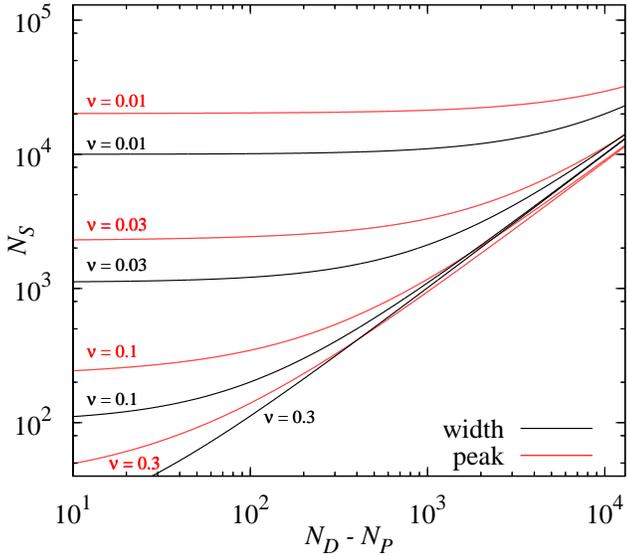}
\vspace{-0.1cm}
\caption{Same as Figure \ref{fig:fom}, but for the fractional error, $\nu$, from the variance on the unbiased parameter covariance matrix estimated from the unbiased width estimator (black) and the unbiased, optimal peak estimator (red) as a function of the number of realisations used to estimate the data covariance matrix, $N_S$, and the size of the data vector minus the number of parameters fitted, $N_D-N_P$. The black (red) lines  show  different values of constant error for the width (peak) parameter covariance estimator. In this case both the width and optimal peak parameter covariance estimate only depends on the fractional error, $\nu$, and $N_D-N_P$. }
\label{fig:covnoise_norm}
\end{figure}

Since we now know the bias in the estimated parameter covariance matrix from equation (\ref{eq:DCC}) 
we can form an unbiased width parameter covariance estimator by dividing by the prefactor  so that $\lgl \calC_{\alpha\beta}^{W,U} \rgl  = \calC_{\alpha\beta}$ (note that we have included the correction for the bias in the precision matrix, although this is now unnecessary). The variance of the elements of the unbiased  parameter covariance is
 \be
 	\sigma^2 \left[\calC^{W,U}_{\alpha\beta}\right]  = \frac{1}{N_S-N_D+N_P-1}
	\left(  |\calC_{\alpha\beta}|^2 + \calC_{\alpha\alpha} \calC_{\beta\beta} \right),
	\label{eq:widthcovunbiased}
 \ee
which is simplified.  In the limit $N_S \gg N_D - N_P$ the first order correction to the fractional variance is 
\be
	\nu^2 = \frac{2}{N_S}+2\frac{(N_D-N_P+1)}{N_S^2},
	\label{eq:v_width}
\ee
where the first term is the usual scaling to the error on independent realisations, while the second term adds the effect of the Wishart sampling to the data covariance matrix, and parameter covariance.
Figure \ref{fig:covnoise_norm} shows contours of constant $\nu$ for the error on the unbiased estimator, where $N_S = N_D-N_P+1+2/\nu^2 $. Compared to the scaling for the biased case, equation (\ref{eq:NSwidth}),  the main effect of rescaling is to reduce the dependence on the number of parameters, $N_P$.  For large values we again find $N_S = N_D$.

It is worth noting that our debiasing scheme is exact only for Gaussian-distributed data and Gaussian-distributed parameter posteriors. While the results for the bias and scatter in the precision matrix for non-Gaussian data, and in particular cosmological data, has been demonstrated (Hartlap et al. 2007, Dodelson \& Schneider 2013), the assumption of Gaussian-distributed parameters has not yet been demonstrated on realistic parameter estimation.

\subsection{Parameter covariance from  peak scatter}

\begin{figure}
\centering
\includegraphics[scale=1.]{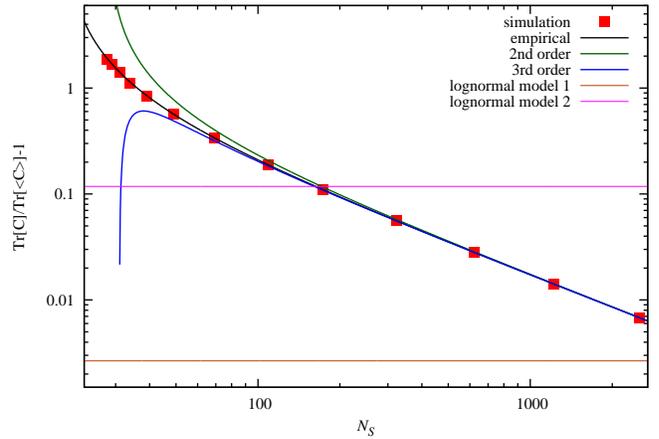}
\caption{The same as Figure \ref{fig:parcov_curv_nd-24}, but for the parameter covariance measured from the scatter in the likelihood peaks.  The green  and blue solid lines shows the second order (Dodelson \& Schneider 2013) and third order solutions. Horizontal lines are for the lognormal Model 1 (orange) and Model 2 (purple).}
\label{fig:parcov_peak_nd-24}
\end{figure}

It appears more difficult to derive the full distribution for the parameter covariance matrix derived from the scatter in the peak of the likelihood. However, we can propose a highly accurate ansatz for the expectation value of the parameter covariance matrix,
\be
	\lgl \widehat{\calC}^P_{\alpha\beta}  \rgl = \frac{N_S-2}{N_S-N_D+N_P-2} \, \calC_{\alpha\beta}.
		\label{eq:cov}
\ee
To second order in $N_S-N_D$, and assuming $N_S-N_D \gg N_P-2$, this yields 
$
	\lgl \Delta{\calC}^P_{\alpha\beta} \rgl \approx [(N_D-N_P)/(N_S-N_D)] \calC_{\alpha\beta}.
$
 which agrees with the second order solution of  Dodelson \& Schneider (2013). In the limit that $N_S \gg N_D \gg N_P$ this further reduces to $\lgl \Delta{\calC}^P_{\alpha\beta} \rgl = (N_D/N_S) \calC_{\alpha\beta}$.
Our expression also agrees with a third-order expansion of equation  (\ref{eq:exact_cov}), where
 $
  	\lgl \Delta{\calC}^P_{\alpha\beta}  \rgl \approx 
	(N_D-N_P)/(N_S-N_D)
	 \left[1- N_P/(N_S-N_D) \right]\, \calC_{\alpha\beta}.
	\label{eq:DCP}
 $
 Both the second and third-order approximations, and by design our ansatz,  reduce to  $\lgl \widehat{\calC}^P_{\alpha\beta}  \rgl = \calC_{\alpha\beta}$ when $N_D=N_P$, due to cancellation of curvature and peak scatter terms in equation (\ref{eq:exact_cov}), as does the width parameter covariance estimate when the bias correction is not applied to the precision matrix.   Figure \ref{fig:parcov_peak_nd-24} shows a comparison of our ansatz with the results of the Wishart-sampler simulations as well as the second order approximation of Dodelson \& Schneider (2013) and our third-order approximation. At high $N_S$ they all agree as expected, but at low $N_S$ the Dodelson \& Schneider result overestimates the increased covariance, while the third order result overcorrects and diverges badly, even at modest $N_S$. Our ansatz proves so simple and accurate compared to the simulations that we cannot help feel it is exact, but we have been unable to prove it at this time.  In the limit when $N_S \gg N_D > N_P$, the ratio of the width to the peak parameter covariance  is $1-(N_D-2N_P+1)/N_S$, hence the width parameter covariance matrix  is always less biased than the peak parameter covariance in this regime. 
 
 As well as the covariance bias we can also estimate the error on the peak covariance matrix, which to our knowledge has not been considered before. 
 The covariance of the peak parameter covariance is more involved, as  we must also take into account the scatter in the data as well as the scatter in the precision matrix. In Appendix \ref{app:covcov} we derive the full expression for the covariance of the deviation in the peak estimator due to a change in the precision matrix, taking the expectation over realisations of the data (equation \ref{eq:devcovcov}). 
 
 Since we do not know the distribution of the peak parameter covariance we again fit for the covariance. We have some guidance on its possible form based on our result for the width parameter covariance. However we have little guidance from a series expansion which to lowest order  is fourth order in the precision matrix and yields 105 terms when averaged. We have already seen that the third order  expansion for the peak parameter covariance diverges badly and so we do not pursue a series solution further.

\begin{figure}
\centering
\includegraphics[scale=1.]{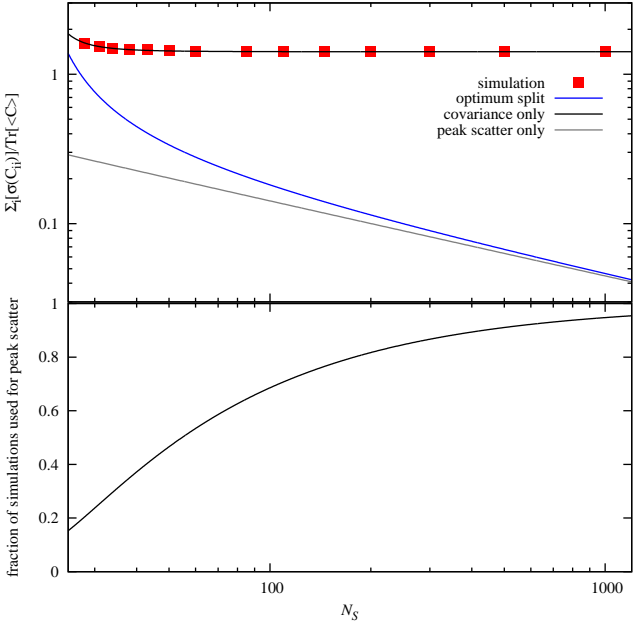}
\caption{\textit{Top panel}: Sum of the standard deviations of the diagonal elements of the unbiased parameter covariance matrix measured from the scatter in the likelihood peaks as a function of the number of simulations used to generate the data covariance matrix, $N_S$, for $f=0$, with no pre-factor in equation (\ref{eq:A}). The standard deviation is normalised by the trace of the noise-free parameter covariance. Red squares correspond to Wishart sampled simulation results for the covariance scatter only, the black solid line to the scaling given by equations (\ref{eq:peakcov}) and (\ref{eq:A}), again for $f=0$  and no pre-factor. The grey line is the limit expected for peak scatter only, while the blue line shows the standard deviation if the $N_S$ simulations are optimally split between estimation of covariance and peak scatter. \textit{Bottom panel}: The optimal fraction, $f$, of the $N_S$ simulations used to determine the peak scatter.}
\label{fig:peakopt}

\end{figure}

	Since we already have an accurate expression for the bias on the peak parameter covariance, equation (\ref{eq:cov}), we can debiased the peak covariance estimate. An empirical fit to our Wishart-sampler simulations for the debiased peak covariance estimator is; 
   \be
 	\left\lgl \Delta \calC^{P,U}_{\alpha\beta} \Delta \calC^{P,U}_{\mu\nu} \right\rgl 
	=A \left( \calC_{\alpha\mu} \calC_{\beta\nu} 
	+ \calC_{\alpha\nu} \calC_{\beta\mu} \right),
	\label{eq:peakcov}
 \ee
where the coefficient is accurately given by
\be
		A=  \frac{1}{N_S f-1}\left( 1 +  \frac{1.6(N_D-N_P)^{0.73}}{[N_S (1- f)- 0.87(N_D-N_P)-2]^2} \right)^2,
	\label{eq:A}
\ee
and we have debiased the peak estimator so that $\lgl \widehat{\calC}^{P,U}_{\alpha\beta}\rgl = \calC_{\alpha\beta}$,
and the numerical coefficients and index are measured to a few percent accuracy from the numerical realisations. 
A fraction, $f$, of the realisations are used to estimate the scatter in the peak likelihood, and $1-f$  is the fraction  used to estimate the data covariance and precision matrix, where we have kept the overall number of realisations fixed at  $N_S$. 
The upper panel of Figure~\ref{fig:peakopt} shows the scaling of equation (\ref{eq:A}) without the overall prefactor of $1/(N_S f-1)$, and $f=0$ in the bracketed term, compared to the numerical Wishart-sampler. The agreement is again very good.

We can find the optimal value of $f$ by minimising $A$ to yield a cubic equation which we can solve for $f$.  For  $N_S \gg N_D - N_P$ we find the optimal fraction is well approximated by $f = 1-2.24 (N_D-N_P)^{0.5}N_S^{-0.725}$.
 The lower panel of Figure~\ref{fig:peakopt} shows the optimal  fraction of realisations needed to estimate the peak scatter as a function of number of realisations. For $N_S \approx N_D-N_P$  the fraction of realisations to estimate the data covariance increases towards unity, while for $N_S \gg N_D-N_P$ a small fraction are used to estimate the data covariance and most are used to reduce the scatter in peaks. In this latter regime we can substitute the optimal fraction  back into the expression of $A$ and, taking only leading terms, we find the fractional error for the unbiased, optimal peak estimator is 
 \be
 	\nu^2=\frac{2}{N_S} +4.48\frac{ (N_D-N_P)^{0.5}}{N_S^{1.725}}.
\ee
Compared to the fractional variance on the unbiased width estimator, equation (\ref{eq:v_width}), the optimal peak fractional error  grows slower with $N_D-N_P$ and falls off slightly slower with $N_S$, as shown in Figure \ref{fig:covnoise_norm}. 
In general, as we can see from Figure \ref{fig:covnoise_norm}, the unbiased width parameter covariance estimator requires fewer realisations, $N_S$, to reach a given accuracy than the unbiased peak estimate. Hence, with the caveat that we have assumed Gaussianity in the data and parameter distribution, we advocate the sampled width estimator over the peak estimator.

\begin{figure}
\centering
\includegraphics[scale=0.68]{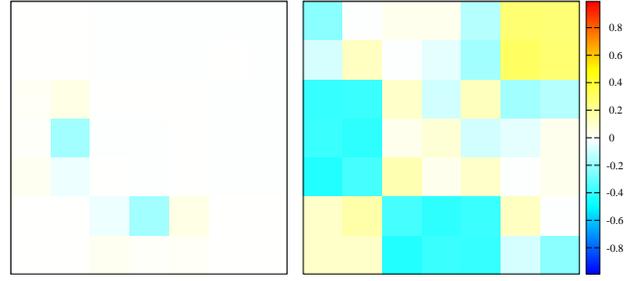}
\caption{Same as Figure \ref{fig:fullmatrix_curv}, but for the parameter covariance matrix measured from the scatter in the likelihood peaks.}
\label{fig:fullmatrix_peak}
\end{figure}

We again  check the accuracy of the off-diagonal terms in the parameter covariance matrix, not least because we have assumed a Wick relation form based on the Wishart-distribution of the width parameter covariance.   Figure \ref{fig:fullmatrix_peak}  shows the relative difference between the our ansatz for the 7-parameter peak covariance matrix and results from our Wishart-sampler. The difference is at the fraction of a percent level ($<0.5\%$) for all terms in the parameter covariance. The right panel in the figure shows the similar factional difference in the variance of the 7-parameter covariance matrix from our empirical fit and the Wishart-sampler. Again the accuracy is sub-percent ($<0.3\%$), which strongly implies that we can write the covariance of the peak parameter covariance with a renormalised Wick-term.

\section{Data compression and modelling}
\label{sec:data_compress}

As both peak and width estimators scale with $N_D-N_P$ we can remove the bias induced by the Wishart scatter, and minimise the additional Wishart covariance, by compressing the data vector to $N_P$ points. This corresponds to  maximal compression, since  further compression must introduce degeneracies between the estimated parameters. Interestingly, our formulae suggest that the same effect may be achieved by inflating the number of parameters. However, as we show in Appendix \ref{app:priors}, this is of no practical use due to the impact of priors which would need to be imposed on these additional parameters.

The simplest, linear compression of the data is a Karhunen-Lo{\`e}ve transformation (e.g., Tegmark, Taylor \& Heavens 1997) which maps  the data to a new vector with the same size as the parameter-space, i.e., $X_\alpha = \A_\alpha^t \widehat{\Psib} \Delta \D$, where $\widehat{\Psib}$ is now a model for the precision matrix. The peak of the likelihood for the compressed data, $X_\alpha$, can be found from  $\Delta \theta_\alpha = - \widehat{\calF}^{-1}_{\alpha\beta} X_\beta$, where $\widehat{\calF}_{\alpha\beta} = \A_\alpha^t \widehat{\Psib} \A_\beta$. The data compression step does not bias the maximum likelihood solution  since  the expectation value of $X_\alpha$ with respect to the data vanishes, $\lgl X_\alpha \rgl =0$. The parameter covariance matrix can be found either by estimating the width of the likelihood in parameter space, $\widehat{\calC}_{\alpha\beta}=\widehat{\calF}_{\alpha\beta}^{-1}$, which adds a fractional statistical error of $\sqrt{2/(N_S-1)}$ from the Gaussian scatter in the realisations of the data (equation \ref{eq:widthcovunbiased}), or from the scatter in the peak likelihood from a set of $N_S$ realisations of the compressed data which also adds a fractional statistical error  of $\sqrt{2/(N_S-1)}$ (equation \ref{eq:peakcov}).    This is  identical to assuming a model for the precision matrix and estimating the parameter values and covariances without a data compression step,  therefore our results apply in both cases.

As  the model for the precision matrix will not be exact, the estimate of  parameter covariance matrix will be biased for both the width and peak estimators. The degree by which the parameter covariance is changed can be estimated from equations (\ref{eq:curv-cov}) and (\ref{eq:exact_cov}), from which  we expect the sensitivity to the model precision matrix is less for the peak covariance. 
If we use the sample precision matrix as our model, we re-introduce the Inverse-Wishart scatter and recover the results for the sampled approach.

As an example of  data compression, or modelling, let us assume  that the covariance of the data is diagonal with a small off-diagonal component, so that $\C = \C_0 (\I + \R)$ where $|R_{ij}| \ll 1$. For a single, linear parameter, $\theta$, where $\mub \propto \theta$,  the parameter variance is
\be
		\calC_{\theta\theta}   =  M_0 \theta^2 [\mub^t (\I+\R)^{-1} \mub]^{-1},
\ee
and where $\hat{\mub}=\mub/|\mub|$ is proportional to the unit data vector.
If our  model  data covariance matrix is proportional to the unit matrix, $\widehat{\C}=M_1 \I$,  the width parameter variance estimator yields
\be
	\widehat{\calC}^W_{\theta\theta} = M_1 \left(\frac{\theta}{\mu}\right)^2 ,
\ee
which misses the covariance terms and will be biased if $M_1$ differs from $M_0$. The peak scatter parameter covariance estimator will yield 
\ba
		\widehat{\calC}^P_{\theta\theta} &=&  [\Tr \, (\A_\alpha \A^t_{\mu})]^{-1}    [\A_\mu^t \C \A_{\nu}]
									 [\Tr \, (\A_\nu \A^t_{\beta})]^{-1} \nn
									 &=&
									  M_0 \left(\frac{\theta}{\mu}\right)^2 \big(1+ \hat{\mub}^t \, \R \, \hat{\mub} \big),
\ea
which reproduces the parameter covariance to first order in $\R$ and has a fractional error  of $\sqrt{2/(N_S-1)}$.
We assume that the realisations of the data used to produce the scatter in the likelihood peak have the same covariances as the data.

\begin{figure}
\centering
\includegraphics[scale=1.3]{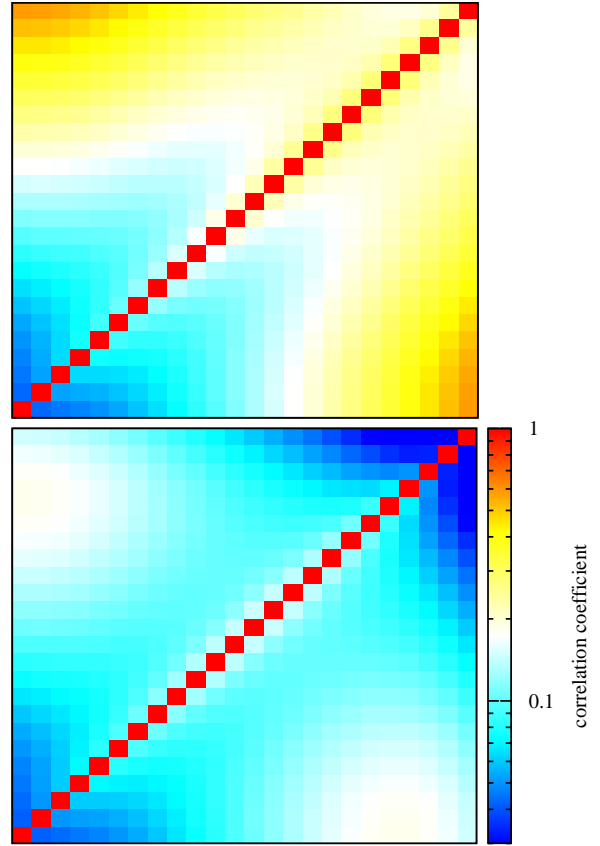}
\caption{\textit{Bottom panel}: Binned correlation matrix, $r_{ij}$ for the lognormal lensing convergence angular power spectrum covariance for Model 1 with shape noise $\sigma_\epsilon=0.35$. As the shape noise is Gaussian, the correlation coefficient is close to diagonal. \textit{Top panel}: Same as bottom panel, but for Model 2 with $\sigma_\epsilon=0$. The stronger non-linearity introduces off-diagonal correlations between power on different scales. }
\label{fig:lognormal}
\end{figure}

A second example is if we have assumed the model is based on Gaussian distributed data, but the real data is non-Gaussian. To explore this we construct a lognormal covariance model which we can compare with the realisations from the Wishart-sampler based on Gaussian data. The extra covariance in the lognormal model, in addition to the Gaussian term,  is given by (see Asgari et al. 2014, for details) 
\ba
M^{\rm LN}_{\ell_1 \ell_2}  \!\!\!\!\! &=&  \!\!\!\!\!\lgl \Delta P_\kappa(\ell_1) \Delta P_\kappa(\ell_2) \rgl_{\rm LN}\nn
&& \hspace{-1.5cm} =	\frac{2}{A_{\rm s}\, \kappa_0^2} \Big[ P^2_\kappa(\ell_1) P_\kappa(\ell_2) +  P_\kappa(\ell_1) P^2_\kappa(\ell_2) + [P_\kappa(\ell_1)+ P_\kappa(\ell_2)]^2 \nn 
& & \hspace{-0.5cm} \times\; \frac{1}{2\pi} \int_0^\pi \!\! {\rm d} \phi\; \big\{ P_\kappa(|\lb_1-\lb_2|)  +	P_\kappa(|\lb_1+\lb_2|) \big\} \Big],
\ea
where $\phi$ is the angle between $\lb_1$ and $\lb_2$,
$P_\kappa(\ell)$ is the lensing convergence power spectrum, $A_s$ is the area of the survey, and $\kappa_0$ is the absolute value of the minimum convergence (see Joachimi et al. 2011). We adopt $\kappa_0=0.012$ from Table~1 of Hilbert et al. (2011) who estimated the minimum convergence value from the Millennium Simulation with weak lensing source galaxies at $z=0.76$. The full model data covariance matrix is $M_{\ell_1 \ell_2}= M^{\rm Gauss}_{\ell_1 \ell_2}+M^{\rm LN}_{\ell_1 \ell_2}$.  
Figure \ref{fig:lognormal} shows the correlation matrix, $r_{ij}=M_{ij}/\sqrt{M_{ii} M_{jj}}$, for the binned lognormal distributed convergence power spectrum. 
The lower panel of Figure \ref{fig:lognormal} (Model 1) assumes a realistic scatter in the source galaxy ellipticity, or shape noise, of $\sigma_\epsilon = 0.35$, showing the effect of Gaussian noise which de-correlates the higher $\ell$-modes, while the upper panel assumes the underlying galaxy population has zero ellipticity which shows the full effect of non-Gaussianity (Model 2) where the higher $\ell$-modes are more correlated.

The lower panel in Figure \ref{fig:parcov_curv_nd-24}  shows the bias on the width parameter covariance estimator from Model 1 (lower line, realistic shape noise)  and 2 (upper line, no shape noise).  Lognormal Model 1, with shape noise, introduces just over a $10\%$ extra bias in the width parameter covariance. Compared to the random sampling approach we would improve on this with $N_S >85$ realisations, for $N_D=29$ and $N_P=7$. For the lognormal Model 2, without shape noise, the fractional bias in the width parameter covariance is $100\%$, which could be improved upon with sampling only $N_S> 30$ realisations. 
Figure \ref{fig:parcov_peak_nd-24} shows the same, but for the peak scatter parameter covariance estimate. Lognormal Model 1, with shape noise, generates only $\approx 0.3$ percent bias in the parameter covariance estimate due to the low sensitivity of the peak estimator, and beats the sampling approach unless we use many thousands of realisations. Model 2, with no shape noise, generates a $10\%$ bias which can be beaten by the sampling approach with $N_S>200$. In this example the less-sensitive model peak parameter covariance estimator is an order of magnitude  less biased than the width estimator.

In general, we expect the total fractional uncertainty on a modelled, or data compressed,  estimator is  
\be
	\nu_X^2 = \frac{2}{N_S-1} + b_X^2
\ee
where $b_X = \Delta C_{\alpha\alpha}/C_{\alpha\alpha}$  and $\nu_X$ are the fractional bias and error on the width ($X=W$) and peak scatter ($X=P$) parameter covariance due to the model, added in quadrature. 
We can compare this with the sampled width estimator, where the fraction variance was given by equation (\ref{eq:widthcov}). For fixed total fractional variance, the model estimators requires fewer independent realisations when 
\be
	\frac{b_X^2}{\nu^2} < \frac{(N_D-N_P)}{(N_D-N_P)+2/\nu^2},
\ee
where we assume $\nu_X=\nu$.
For $(N_D-N_P) \gg 2/\nu^2$ we require $b_X < \nu$ or else the accuracy cannot be met, while for $(N_D-N_P) \ll 2/\nu^2$ we need $b_X< \nu^2 \sqrt{(N_D-N_P)/2}$. The modelled peak estimator will perform better than the width estimator as the bias is lower, in our examples an order of magnitude better. Assuming we want at least $10\%$ errors on parameters, for $N_D-N_P \gg  200$ , we need the modelled peak bias to be better than $10\%$, while for $N_D-N_P <  200$ we need $b_X< 1.4 \times 10^{-2} \sqrt{(N_D-N_P)}$.

To reach firmer conclusions on the promise of analytical covariance models, the toy comparison between a lognormal model and a Gaussian covariance will have to be replaced with a confrontation of covariances extracted from suites of N-body simulations against realistic models. These could for instance be based on the assumption of lognormality (see Hilbert et al. 2011), or built via the halo model (Pielorz et al. 2010). Shrinkage estimation (e.g. Pope \& Szapudi 2008) allows for a smooth transition between a pure model-based covariance and a pure sample covariance, automatically balancing bias and variance. This will be investigated in a forthcoming paper.

\section{Summary and Conclusions}
 
\begin{table*}
\centering
\caption{Summary  of new results for the statistics of parameter covariances presented in this paper. Equation numbers refer to results in this paper, following citations refer to previous second-order solutions.}
\begin{tabular}[t]{lll}
\hline\hline
 & Width parameter covariance & Peak parameter covariance\\
\hline
Bias in the mean estimator & Eq. (\ref{eq:DCC}) (2nd order Percival et al. 2013) &  Eq. (\ref{eq:cov})  (2nd order Dodelson \& Schneider 2013)\\
Variance of biased estimators &  Eq. (\ref{eq:widthcov})  (2nd order Taylor et al. 2013) &     \\
Variance of unbiased estimators & Eq. (\ref{eq:widthcovunbiased}) & Eq. (\ref{eq:peakcov})\\
\hline
\end{tabular}
\label{tab:structure}
\end{table*}

As Cosmology progresses into the age of high precision measurements, probing the nature of dark energy and dark matter, and looking for evidence of modified gravity,  accurate estimates of cosmological parameter errors and covariances require accurate data covariance matrices.
Due to the nonlinearity in cosmological fields, and the complexity and nonlinearity of data  analysis pipelines, a common approach is to estimate the data covariance from an ensemble of realisations of the survey, usually using simulations or empirically from the data from Jackknife or Bootstrap resampling. 
In maximum likelihood parameter estimation, errors and covariances  are derived from the width of the likelihood surface or from the scatter in the peak of the likelihood from a set of realisations. However,  statistical uncertainty and bias in the estimation of the data covariance matrix will propagate into additional errors on the measurement of cosmological parameters.

In this paper we have found expressions for the change in the parameter covariance for both width and peak scatter estimators, assuming that the data and parameters are Gaussian-distributed,  due to a change in the data covariance matrix from its expectation value. In particular, the width error estimate is biased at first order while the peak scatter error estimate has a second order bias, and so the error estimates from both estimators will not in general coincide.

If the data covariance matrix  for $N_D$ data points is estimated from $N_S$ independent random Gaussian-distributed realisations of the data, it will be Wishart-distributed with $N_S-1$ degrees of freedom. We have shown here that the parameter covariance matrix for $N_P$ parameters,  estimated from the width of the likelihood, is also Wishart-distributed with $N_S-N_D+N_P-2$ degrees of freedom. With the full distribution of the parameter covariance matrix we have derived its  expectation value (equation \ref{eq:DCC}) and its covariance (equation \ref{eq:widthcov}). In general, the estimated parameter covariance will be larger than the optimal parameter covariance matrix, estimated from an infinite number of realisations, but we can renormalise the estimated parameter covariance matrix to find an unbiased estimate. The fractional variance of the unbiased width estimator then scales as $\nu^2 = 2/(N_S-N_D+N_P-1)$.
Using our expressions for the change in the parameter covariance due to the change in the data covariance, and a Wishart sampler to generate realisations of the data covariance matrix for $N_S$ realisations, we have numerically estimated the expectation and variance of the parameter covariance and found excellent agreement with our analytic results.
Table \ref{tab:structure} summarises the key expressions  for biases and variances of the parameter covariance matrices.

The distribution of the peak scatter parameter covariance matrix, given a Wishart-distributed data covariance, does not appear tractable but we have found a very accurate ansatz for its expectation value (equation \ref{eq:cov}). Again we have used this expectation value, which is again larger than the optimal estimate, to renormalise and find an unbiased peak scatter estimator. The covariance of the parameter covariance matrix is more complex, as we need to have independent realisations of the data to estimate both the data covariance matrix and to reduce the uncertainty in the estimate arising from the random scatter in the position of the likelihood peak. For a fixed number of realisations, we find the optimal split between the data covariance and peak scatter estimates and fit an accurate expression for the covariance of the unbiased estimate (equation \ref{eq:peakcov}). 
We have found that the parameter covariance estimated from the peak scatter is generally larger than the width estimate. For the unbiased estimators the variance of the optimal peak scatter estimator is also larger than that of the width estimator, and so in general we advocate the width estimator over the peak scatter estimator.

The Wishart bias for both peak and width estimators vanishes if the number of data points equals the number of parameters, however the  scatter on both  scales as $\sqrt{2/N_S-1}$. 
We can achieve $N_D=N_P$ by data compression, since expanding the parameter space does not help. 
In our analysis data compression is equivalent to assuming a model for the data covariance matrix, and the accuracy in the parameter covariance in both cases  is determined by the accuracy of the model. 
The width parameter covariance estimator is more sensitive to biases in the data covariance model than the second order peak scatter  estimator to the assumed data covariance.This implies that for data compression, or modelling, the model data covariance matrix will be more accurate for the peak estimator than the width estimator 
The peak scatter estimator requires independent realisations to estimate the parameter covariance and we need  $N_S > 1+2/(\nu^2-b_P^2)  $ realisations to reach a fractional accuracy of $\nu$ in the parameter covariance matrix, where $b_P$ is the fractional  bias from the model data covariance matrix.

 In current and future cosmological surveys where the number of data points, for example power spectra bandpass modes or correlation function points in redshift bins, will grow from hundreds to thousands or tens of thousands,  to control the accuracy of errors we will have to generate $N_S > N_D-N_P+2/\nu^2+1$ independent realisations of the data, or accurately model the data covariance matrix. If we sample independent realisations to estimate the data covariance, we have shown that we can remove the inherent bias in estimates of the parameter covariance matrix and control its error, preferring a width estimator over a peak scatter estimator. If we model the data covariance matrix, or apply data compression, we find the lower sensitivity of the peak scatter estimator makes it preferable to the width estimator, but the peak scatter may need large numbers of  independent realisations to reduce the random error on the parameter covariance if the bias is significant. Further work is required to assess if the model bias is low enough to apply the peak scatter estimator. If not, the sampled unbiased width estimator is preferred.

\section*{Acknowledgements}
 We thank John Peacock for stimulating discussions and in particular  drawing our attention to the potential of modelled peak estimators, and  Tom Kitching, Alina Kiessling, Scott Dodelson and Tim Eifler for useful discussion during the development of this paper. We also thank the Aspen Centre for Physics who hosted the excellent Workshop on Weak Lensing Surveys where this paper began. BJ acknowledges support by an STFC Ernest Rutherford Fellowship, grant reference ST/J004421/1.

{}

\appendix

\section{Covariance of the peak estimator}
\label{app:covcov}

In this Appendix we derive the full parameter covariance matrix covariance for the peak estimator, taking into account scatter in both the data and a change in the precision matrix from its expectation. To derive this we start with the deviation in the parameter covariance matrix due to a change in the precision matrix, and prior to averaging over the data, is
\be
	\Delta \calC^P_{\alpha\beta} =  \widehat{\calF}^{-1}_{\alpha \alpha'}( \widehat{d}_{\alpha'} \widehat{d}_{\beta'}  )  \widehat{\calF}^{-1}_{\beta'\beta} - \calF^{-1}_{\alpha\beta},
\ee
 which can be re-written as
 \be
	\Delta \calC^P_{\alpha\beta} =
	\left[ \widehat{\calF}^{-1} \left( \widehat{d} \widehat{d}  - \calF + 2 \Delta \calF + \Delta \calF \calF^{-1} \Delta \calF
	\right) \widehat{\calF}^{-1} 
	\right]_{\alpha\beta}\!\!.
	\label{eq:devcov}
\ee
The expectation value of $ \widehat{d} \widehat{d}$ with respect to the data is
\be
	\lgl  \widehat{d} \widehat{d} \rgl = \calF + 2 \Delta \calF + \Delta \Delta \calF,
\ee
which, when substituted into equation (\ref{eq:devcov}), leads to the resulting equation (\ref{eq:exact_cov}) for the expectation of the deviation in the parameter covariance over the data. To find the covariance of the deviation in the parameter covariance we take equation (\ref{eq:devcov}) and, after taking the expectation over the data, we find the covariance of the deviation is 
\be
\!	\lgl  \Delta \calC^P_{\alpha\beta}  \Delta \calC^P_{\gamma\delta } \rgl  
	\!=\!
	\frac{1}{N_S'-1} \! \left(\widehat{\calC}^P_{\alpha\gamma}\widehat{\calC}^P_{\beta\delta}
	\!+\! \widehat{\calC}^P_{\alpha \delta}\widehat{\calC}^P_{\beta \gamma}
	\!+\!
	 \Delta \calC^P_{\alpha\beta}  \Delta \calC^P_{\gamma\delta } \right)
	 \label{eq:devcovcov}
\ee	
where
\be
	 \widehat{\calC}^P_{\alpha\beta} =\left[ \widehat{\calF}^{-1} \left(\calF + 2 \Delta \calF + \Delta \Delta \calF \right)\widehat{\calF}^{-1}\right]_{\alpha \beta}
\ee	
and
\be
 \Delta \calC^P_{\alpha\beta} = \left[ \widehat{\calF}^{-1} \left( \Delta \Delta \calF -  \Delta \calF \calF^{-1} \Delta \calF\right)\widehat{\calF}^{-1}\right]_{\alpha \beta},
\ee
and $N_S'$ is a set of independent realisations not used to estimate the precision matrix. The first two terms in equation (\ref{eq:devcovcov}) are the usual Wick result for the covariance of Gaussian-distributed data, while the third term is an extra covariance introduced by the change in the precision matrix. If the deviation in the precision matrix, and hence Fisher matrices, vanishes, this reduces to
 \be
	\lgl  \Delta \calC^P_{\alpha\beta}  \Delta \calC^P_{\gamma\delta } \rgl  
	=
	\frac{1}{N_S'-1} \left({\calC}_{\alpha\gamma}{\calC}_{\beta\delta}
	+{\calC}_{\alpha \delta}{\calC}_{\beta \gamma} \right).
\ee

\section{Effect of parameter priors}
\label{app:priors}

\begin{figure}
\centering
\includegraphics[scale=1.]{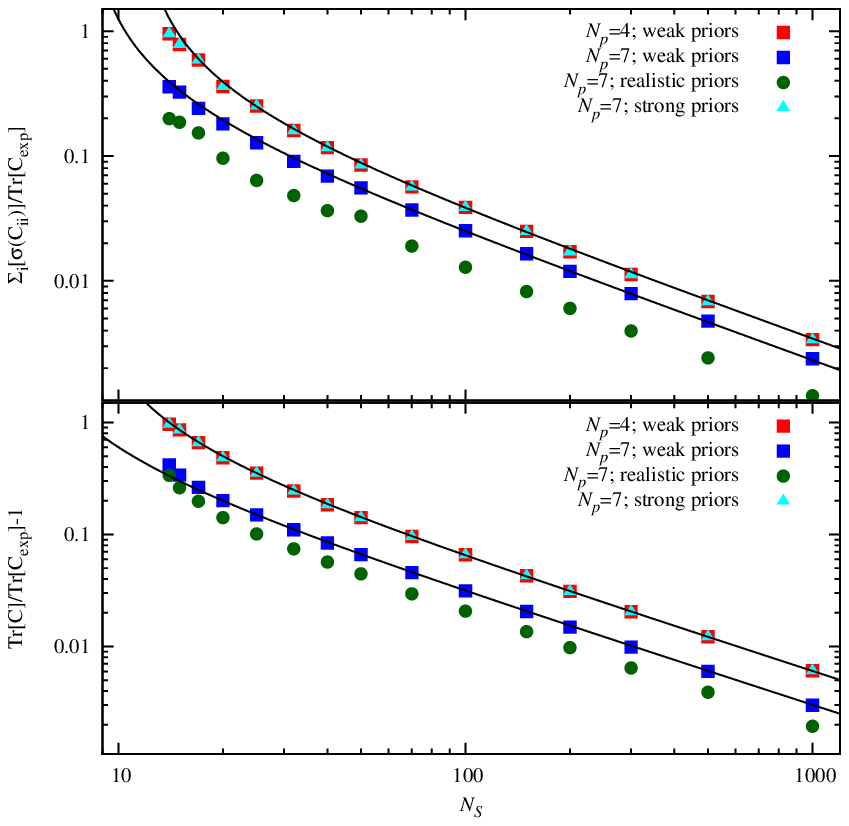}
\caption{Impact of priors on the bias and scatter in the width parameter covariance. \textit{Top panel}: Sum of the standard deviations of the diagonal elements of the parameter covariance as a function of the number of simulations used to generate the data covariance, $N_S$. Red squares correspond to simulation results for $N_P=4$ parameters, using weak priors ($\sigma_p=100$). Blue squares (green circles, cyan triangles) show results for $N_P=7$ and weak (realistic [$\sigma_p=\{0.01,0.5,5\}$ for $\{\Omega_{\rm b},w_0,w_a\}$, respectively], strong [$\sigma_p=10^{-3}$]) priors. The black solid lines are the analytic scaling for $N_P=4$ and $N_P=7$, respectively. \textit{Bottom panel}: Same as above, but for the mean of the trace of the parameter covariance.}
\label{fig:prior}
\end{figure} 

In most applications informative priors will be applied to cosmological parameters as well as nuisance parameters. Gaussian priors generate an additive term to the Fisher matrix and thus affect the parameter covariance in a non-linear way. We use the Wishart sampler to test the impact of priors on three (${\Omega_{\rm b},w_0,w_a}$) of the $N_P=7$ cosmological parameters, showing the bias and variance on the width parameter covariance in Figure \ref{fig:prior}. Wide priors (width of the Gaussian prior $\sigma_p=100$) do not affect results, whereas tight priors ($\sigma_p=10^{-3}$) increase the bias and variance to the level expected for $N_P=4$ parameters.

This finding demonstrates that it is impractical to inflate the number of parameters to $N_D$ with the purpose of eliminating bias and scatter in the parameter covariance. To avoid significant impact on cosmological constraints, one would impose tight priors on any extra parameters introduced into the analysis, and this brings the bias and scatter in the parameter covariance back to its original level.

Finally, we implement an intermediate case for which the order of magnitude for the priors has been matched to current knowledge, setting $\sigma_p=\{0.01,0.5,5\}$ for $\{\Omega_{\rm b},w_0,w_a\}$, respectively. In this case the bias and variance of the parameter covariance lie below both the $N_P=4$ and $N_P=7$ results with weak priors. The prior information suppresses the impact of noise in the parameter covariance while leaving sufficient \lq wiggle-room\rq\ for the parameters to avoid an effective reduction in the dimension of parameter space, which would again boost scatter and bias. Consequently, the scalings derived in this work can be considered conservative, as long as any parameters with tight priors are not included in the count towards $N_P$.

\end{document}